# A Vehicle for Research: Using Street Sweepers to Explore the Landscape of Environmental Community Action

Paul M. Aoki,[1] R.J. Honicky,[1,2] Alan Mainwaring,[1] Chris Myers,[1] Eric Paulos,[1,3]
Sushmita Subramanian,[1] and Allison Woodruff [1]

| [1] Intel Research | [2] EECS Department | [3] Carnegie Mellon Univ. |
|---|---|---|
| 2150 Shattuck Ave., Ste. 1300 | Univ. of California, Berkeley | 5000 Forbes Ave. |
| Berkeley, CA 94704-1347 USA | Berkeley, CA 94720-1776 USA | Pittsburgh, PA 15213-3891 USA |

**ABSTRACT**
Researchers are developing mobile sensing platforms to facilitate public awareness of environmental conditions. However, turning such awareness into practical community action and political change requires more than just collecting and presenting data. To inform research on mobile environmental sensing, we conducted design fieldwork with government, private, and public interest stakeholders. In parallel, we built an environmental air quality sensing system and deployed it on street sweeping vehicles in a major U.S. city; this served as a "research vehicle" by grounding our interviews and affording us status as environmental action researchers. In this paper, we present a qualitative analysis of the landscape of environmental action, focusing on insights that will help researchers frame meaningful technological interventions.

**ACM Classification Keywords**
H5.m. Information interfaces and presentation (e.g., HCI): Miscellaneous.

**INTRODUCTION**
Reflecting a resurgence of popular concern about environmental sustainability, the HCI community has recently been searching for ways in which its abilities and disciplinary concerns can be brought to bear on environmental issues. However, such societal-scale issues are canonical wicked problems, ones in which stakeholders have radically different views such that the definition and explanation of a problem, the formulation and acceptability of possible solutions, and the meaning and permanence of success are broadly contested [29]. Rather than taking on these problems directly, current HCI sustainability research typically formulates them in classic HCI terms, proposing "green" interventions at the level of personal behavior modification or individual product design; such formulations embody very specific assumptions about a problem, ones that implicitly rely on the logic of market preferences (expressed or revealed) for scale [10]. By contrast, the clear consensus of a CHI 2008 panel on "HCI & Sustainability" was that the field's research focus ought to be on ways to effect systemic, collective change [26].

Here, we consider opportunities, challenges and considerations for the HCI community in developing technology to facilitate environmental change via political processes. We do so in the context of a specific technology that is often motivated in terms of environmental concerns: mobile participatory sensing, in which everyday citizens use sensor-equipped mobile devices to collect and share politically relevant data such as air quality measurements [3]. Mobile sensing is a very active technical research area, particularly for systems researchers (e.g., [3,12,21]). However, little is known of how such systems might fit into the context of real-world environmental action or how diverse stakeholders might generate and make sense of the data they produce. In order to inform future applications of mobile and pervasive technology, we conducted design fieldwork on the social and organizational landscape of environmental action – government agencies, public health NGOs, atmospheric scientists, and so on.

In this paper, we report results from this investigation. Our primary contribution is a qualitative analysis of the landscape of environmental action for air quality, focusing on insights that will help researchers frame meaningful and effective interventions. For example, we describe the various stakeholder perspectives in order to help researchers interact effectively with different parties and to illuminate the context in which technologies and data will be received – or, as we shall see, be judged as irrelevant. We also discuss design implications for HCI, such as the need for social mapping tools that help environmental advocacy groups connect with each other in order to establish important relationships and gain access to critical resources. The specifics we report here are U.S.-centric, but we believe that the analysis will broadly benefit HCI and systems researchers seeking to enter the environmental action domain as well as researchers already working in it. We further hope that this work will serve as an illustrative example of how HCI can engage with political processes.

As a secondary contribution, we describe our deployment of mobile air quality sensing platforms on the municipal fleet of street sweeping trucks in San Francisco, California, and its uses in advancing our research program. We focus here on the role the deployment played in the execution of our fieldwork: serving as a "research vehicle" by affording us status as environmental action researchers in our interactions with governmental and non-governmental actors alike.

The paper is organized as follows. We first describe the context and methods of our investigation. We then analyze the landscape of environmental action. Next, we describe critical perspectives on data interventions such as ours. Finally, we discuss pragmatic considerations and implications for research agendas for environmental action.

**CONTEXT AND METHODS**
Our goal in this study was to understand where information and communication technology (ICT) interventions could play a role in environmental decision-making. From the literature, it was immediately obvious that ICT could play a facilitating role in both internal and external communication – for example, in the case of non-governmental organizations (NGOs), relevant ICT applications include public websites, online communities, fund-raising and outreach campaigns, etc. [14,24]. It was less clear where ICT could play a substantive role in improving decision-making – for example, where new types of data are perceived as having influence, and by whom.

This study considers two main questions: first, where the U.S. environmental decision-making system affords opportunities for "outsider" technological intervention to have influence, and second, understanding how proposed interventions might be viewed by various stakeholders. We aim to provide practical guidance for research intervention in the context of the existing system. We do not justify this by arguing that the present system is entirely effective, let alone ideal or just; while it has had its specific successes with respect to air quality [30], the literature contains many sharp critiques of the U.S. environmental regulatory system [32].[†] Instead, we take the view that it is helpful to understand what (relatively) immediate steps might be taken because there are many documented examples in which local (albeit usually non-technological) interventions have had success in the past [2,5,32] and because this is a realistic scope for action in the context of HCI research.

**Related Work**
There is a vast literature on environmental policy, drawing from disciplines such as political science and public policy (e.g., [2,32]), social studies of science (e.g., [19,20]), environmental sociology (e.g., [4,9]), and urban planning and public health (e.g., [6]). However, as is frequently pointed out [4], field research in this area typically focuses on a single organizational actor (e.g., an ethnography of a specific social movement organization such as Greenpeace) or on a class of such actors (e.g., a historical analysis of regulatory agencies such as the U.S. Environmental Protection Agency (EPA)). Importantly for technologists wishing to design interventions, the viewpoints and interactions of many different actors within a single context are rarely considered. Similarly, while there are many studies of interventions by environmental activist organizations (e.g., [2,6]), there are no field studies that consider the role of novel technological interventions.

In the computing research literature, the problem of connecting technological innovations with environmental policy and decision-making remains underexplored, though many individual related topics have been examined. For example, recent HCI field research has examined "green" attitudes and practices relevant to ICT consumption [15,18,36]. However, there have not been corresponding detailed studies of environmental issues such as air quality outside of this consumer focus. As another example, persuasive technology is being actively explored in the sustainability domain [13,22]. Focusing on personal behavior, these are complementary to the work here. A third example is mobile participatory sensing. This proposes the use of consumer electronics (such as mobile phones) to capture, process, and disseminate sensor data and to thereby "fill in the gaps" where people go but fixed sensor infrastructure has not been installed. While several groups have connected their sensing software platforms to commercial air quality sensor units mounted on vehicles such as bicycles [12,21] or taxis [28], the question of connecting the results of these technical experiments to social action remains unexamined.

Artists have been very active in directly connecting technology and environmental action. One tactic for building community environmental awareness is to deploy air quality sensors on provocative platforms such as pigeons (www.pigeonblog.mapyourcity.net) or robotic dogs (www.nyu.edu/projects/xdesign/feralrobots). A variant tactic is to apply a cheerful, do-it-yourself ethos to air quality monitoring (www.blackcloud.org). The longer-term problem that follows awareness-building – moving beyond short data collection "campaigns" or art show installations to the mobilization of social action and practical engagement with the environmental decision-making process – generally remains beyond the scope of art projects. As has been recently noted, this risks making community members highly aware of problems without making it equally clear how to address them [8].

---

[†] We do not have space to address open-ended questions such as the role of science in environmental decision-making processes [20]; the efficacy of various means of achieving environmental goals (e.g., "command and control" regulation vs. incentives [4,32]); what constitutes a "natural" environment [7] and, by implication, what kind of environment citizens can justly demand; the long-term aims of environmental action in terms of societal structure (e.g., the "deep green" vs. "bright green" debate in sustainability [36]); or the morality of different conceptual frameworks for balancing interests in environmental policy-making (e.g., economics vs. social justice [27,32,34]) – to give just a few examples.

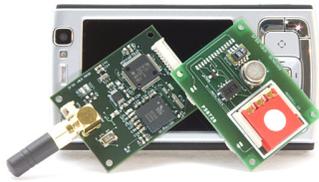 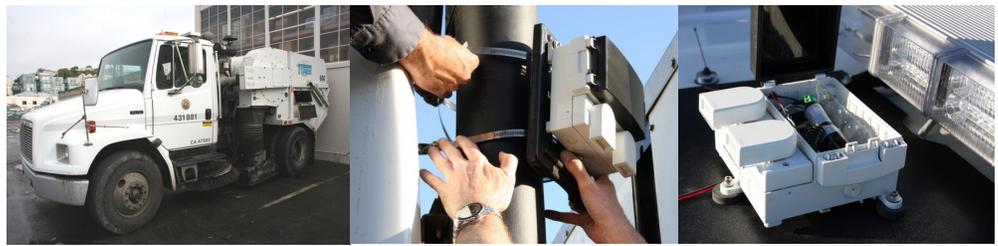

**Figure 1.** Sensor board design.    **Figure 2.** Enclosure-mounting the sensor package on city street sweeping vehicles.

**Intervention**

We conducted this study in the context of the Common Sense project (citizensensing.org), which aims to connect sensing to practical action. Like many of the projects above, Common Sense is developing hardware/software sensing platforms that allow groups and individuals to collect environmental information. Common Sense extends prior research through a focus on collaborative software – mobile and Internet-based software applications that directly support citizens' collective efforts to use environmental information to influence regulations and policy – and on extended deployments in direct collaboration with different types of environmental organizations.

If the kind of sensing-based applications we envision prove out, sensors would be integrated directly into commodity mobile devices. For prototyping, however, we are developing a suite of board designs and embedded software that can be deployed with associated mobile devices or in a stand-alone configuration. The current boards can be selectively populated with commercial carbon monoxide, nitrogen oxides, and ozone gas sensors as well as temperature, relative humidity, and motion sensors (Figure 1). Sensor readings plus GPS data are sent to a database server via GSM text messages. We are also developing mobile and Internet-based visualization tools and community features to support collaborative online interpretation of interesting phenomena (c.f. [16]) and collective development of strategies for practical action.

As an initial technological intervention, we are collaborating with the City of San Francisco to install our air quality sensing systems on the municipal fleet of street sweepers. Street sweepers are vehicles that use mechanisms such as water sprays, brooms, and collection bins to clean debris from city streets (Figure 2). As the vehicles do their work, our devices collect street-by-street air quality readings, the associated mobile phones send the data to our servers, and the data is then displayed in a Web-based application (Figure 3). The street sweeper deployment allows us to leverage mobile city infrastructure, such that a small number of vehicles provides extensive and systematic coverage of a large city, and it gives us an opportunity to test and refine our system in a highly challenging real-world environment, addressing issues such as calibration and possible emissions from the vehicles themselves. Of key relevance to this paper is the role that the deployment serves as a research tool, giving us valuable experience interacting with organizational actors from all parts of the environmental policy landscape.

**Method**

In order to learn about the environmental decision-making process, we conducted fieldwork in the San Francisco Bay Area over a period of seven months in 2008. The Bay Area has specific strategic relevance as a field site for this study. First, since the 1967 Clean Air Act, California has acted as a national air quality "laboratory" [30], authorized to manage its own standards and measures; the California Air Resources Board (ARB) and its 35 air quality management districts manage an annual budget of $750 million, and ARB develops regulatory frameworks around issues such as environmental justice that are often subsequently adopted by other state and federal agencies. Second, there is a reason for this unique role – California has a history of many decades of air pollution troubles [30,33]. Third, the Bay Area has a long history of environmental activism [33], leading to relatively refined stakeholder views.

We conducted formal in-person interviews with 14 stakeholders (active participants who affect and/or are affected by the outcomes of environmental decisions) and informal phone and in-person interviews with approximately 30 more stakeholders. We also interacted with additional stakeholders through email and by visiting worksites and attending meetings. For example, we visited a monitoring station operated by the Bay Area Air Quality Management District (BAAQMD, colloquially known as the "Air District"), we attended community town-hall and activist meetings, we networked at a regional air quality awards event, and we travelled to a national air quality conference where we met with government representatives from the U.S. EPA and several states. We also collected public outreach documents. Finally, we held a community workshop for approximately 20 people.

We used an organic recruiting process that leveraged a combination of resources, e.g., contacts in city government and citizen groups, contacts made at events and meetings, email lists, cold-calling, and assistance from an NGO. The stakeholders we spoke with represented many different perspectives, e.g., city and state government representatives, advisory board members, remediation consultants, air quality consultants, urban planners, physicians, scientists, NGO organizers and volunteers, lung cancer survivors, and many others. Almost all participants were adults, at a variety of life stages, with a fairly balanced number of male and female participants.

The formal interviews were semi-structured and lasted approximately 1.5 to 3 hours. The informal interactions followed a more open-ended format and varied greatly in length. In both the formal and informal interactions, we grounded our discussions with the street sweeper deployment as well as an upcoming deployment of personal sensing devices, for example often soliciting feedback on prototypes and/or showing visualizations of data (e.g., Figure 3). This allowed us to iteratively refine our designs, but more importantly, the deployments made the discussions more concrete and the grounded examples allowed us to explore more deeply people's responses to the properties of mobile environmental sensing. We took detailed field notes on all interactions and we recorded the formal interviews, transcribing relevant segments. We performed an affinity clustering on the textual corpus to identify emergent themes, as well as constructing visual diagrams of how the various parties conceive of and influence each other [1].

## THE LANDSCAPE OF AIR QUALITY MANAGEMENT

Air quality is a high stakes, hotly contested political topic. Airborne pollutants can have both short-term and long-term adverse health effects on the general population, and they can be particularly damaging to children, outdoor athletes, or individuals with respiratory conditions such as asthma. Further, air quality is linked to environmental justice concerns about disproportionate exposure and shorter life spans for disadvantaged populations who live in less desirable areas near industrial facilities, highways, and other hazardous sources [32]. At the same time, links between specific pollution sources and health are often difficult to establish, remediation is often complicated and costly, and parties often have competing interests. Therefore, many public and private parties take an active role in air quality (and, more broadly, environmental) decision-making. In this section, we give an overview of these parties, how they interact with each other, and participants' views on how air quality should be measured. We focus on the points most relevant to data interventions.

### Organizational Roles and Individual Perspectives

Many diverse organizations are active in environmental decision-making. These can be roughly grouped into the following categories: government, emitters, and advocates. Each represents a broadly different institutional view.

*Government:* Legislative, executive and judicial bodies occupy key positions in the environmental landscape. The rise of government air quality regulation in the 1960s led to many new agencies and responsibilities at the national, state, and local levels [30,32]. Roles include establishing policies and regulations; measuring and reporting environmental conditions; and assessing and enforcing regulatory compliance. As the regional government agency with primary responsibility for local air quality measurement, the Air District occupies a central position in the Bay Area air quality landscape.

*Emitters:* Private or government entities which operate facilities such as factories, oil refineries, or power plants, or which conduct activities such as construction or transportation, necessarily produce some industrial pollution. Such entities must balance the expense of emission reduction with the legal, ethical, and public relations liabilities associated with their emissions.

*Public interest advocates*: Many different kinds of NGOs advocate for improved air quality, ranging from national NGOs with significant resources and infrastructure (e.g., the Sierra Club) to small community groups of concerned citizens who meet around a kitchen table. A given organization may have a broad agenda such as fighting lung disease, a local agenda such as reducing emissions from a nearby steel plant, or both.

One would expect that the views of individuals would be somewhat aligned with those of their institution due to self-selection [9]. At the same time, one would not expect view to be determined solely by membership; for example, within a single environmental regulatory organization, scientific, bureaucratic and political professional subcultures co-exist and often come in conflict [11]. We will see such differences later in this section.

### The Work of "Managing" Air Quality

In the U.S., environmental regulations and policy are ultimately created and enforced by governments, in a complex process in which both government and non-government parties participate. A strawman, "rational" view of these processes is that they are cost/benefit decisions about specific activities or policies. For example, government or private parties may propose a new activity, such as building a bridge or an oil refinery, or public interests may propose new regulations – motivated, for example, by new scientific information about hazards. The potential health risks, economic costs and benefits, etc. of such actions can be estimated and weighed. The process is therefore framed in terms of evaluation and argumentation, and private and public actors often oppose each other in judicial and/or administrative venues. For example, some of the community activists we observed were in the process of filing a class action lawsuit demanding that a local factory reduce its emissions; meanwhile, the factory's owners were releasing their own health impact reports that demonstrated compliance to regulators.

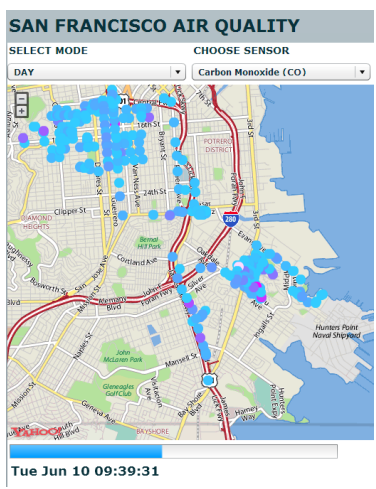

**Figure 3. A sample visualization (detail).**

Still, a view of adversarial relationships between parties in structured venues is clearly oversimplified. Parties attempt to influence each other using pragmatic arguments, political pressure, and media attention as well. Further, advocacy groups, government organizations, and emitters often establish carefully negotiated relationships. As a first example, advocacy NGOs often have close working relationships with legislators and their staffers, providing scientific analyses as well as information about the priorities of their constituencies. Real-world data is useful in drawing attention to problems and advocating change.

> "If [the data shows] there are exposure points along my regular route…I can make the case that this is my daily life and it's being impacted in these ways, and it's then a really strong argument for bringing to a city official." – *Env. Mgmt. student*

A second example is the use of advisory boards; for example, non-government board members provide government organizations recommendations and external expertise on topics such as policy, health impacts, finances, scientific findings, legal issues, and citizen concerns [32]. A final example is the creation of collaborative partnerships among government, advocates, and emitters, in which citizens play an active role in identifying problems, framing research questions, collecting and interpreting data, and suggesting efficient remediation strategies [6].

The constant struggle to influence has direct implications for the way in which advocacy groups organize and operate. Decision-makers at all levels tend to be responsive to larger constituencies. Hence, to maximize influence and to leverage shared efforts, individuals aggregate into groups and groups aggregate into coalitions.

> "You've got to mobilize. Otherwise it's one voice in the wilderness, and they're just going to discount you…At some point, if the voice is loud enough, it will be heard. That's the bottom line." – *NGO volunteer and concerned citizen*

Coalitions do not just increase numbers and mindshare. Established, issue-based organizations such as Breathe California and Communities for a Better Environment offer financial, organizational, and social network resources. Hence, community-based grassroots groups often "hitch" their local agendas to broader issues to gain access to these resources [6].

> "It really came down to Breathe California. When I went to Breathe California…[the director] understood the problem, she also understands the dynamics of the city and everything like that. So she got me in touch with [a member of the council] and she was the most environmentally aware. And then they really fought for us." – *Concerned citizen and activist*

**How Should Air Quality Be Measured?**
The Air District occupies a key position in the Bay Area air quality landscape. Its regulatory mission is to gather high quality data measurements in accordance with federal guidelines and to ensure regional compliance with state and federal standards. However, other parties propose different goals for air quality measurement based on alternative ideas of how to improve public health or design effective remediations. These alternative formulations imply very different needs for data collection and analysis. Differences in opinion regarding the current data collection system largely centered on the limited number of monitoring sites, the location of these sites, and the way the data is applied.

First, participants disagreed whether there are enough Air District monitors to give an accurate picture of air quality throughout the Bay Area. (There is one site in the City of San Francisco and 30 others throughout the Bay Area; in limited cases, additional "temporary" sites are established to monitor specific sources.) Factors such as topography, wind, and temperature are understood to impact the movement and formation of pollutants, and the Bay Area is a large, geographically and meteorologically complex region with many emissions sources. Further, pilot studies commissioned by community groups suggest high variation within given neighborhoods. Consequently, activists often felt there were not enough monitors, as did some atmospheric scientists working for the Air District.

> "The Air District has…twenty or twenty-four, something like that, monitoring points around the region. Okay, so they have good long-term data there, but five blocks from there, you don't have any data, they just assume that it's spread out in a certain way… The more data you get, the better it is. The more different types of data you get the better it is. And it helps improve the models." – *Planner and NGO volunteer*

Second, participants disagreed whether measurements taken by the Air District represent the air that citizens breathe on a day-to-day basis. Both Air District personnel and activists noted that regulations require monitoring intakes to be located high above the ground and well away from highways, railroads, and other identified pollution sources ("hot spots"). This is intended to ensure – per regulatory standards – that monitoring sites collect "representative" values rather than "peak" values. However, activists objected to this approach, pointing out that many citizens are exposed to "peak" conditions in their daily lives. The methods and metrics for assessing health risks are indeed an open scientific question [30,32].

> "We're really interested in figuring out what people breathe at street level… They almost put their sensor in the only place you could be in the neighborhood that [is not] within a thousand feet of a freeway. So, you know, our contention is that if you live here in a condo seventy-five feet from a freeway your level of particulate exposure is much higher than if you live over here… Those are the kind of things we're interested in, and no one so far has been able or willing to try and tell us what those levels are. So that's why this [mobile environmental sensing] is a very exciting concept." – *Community group leader*

Others, such as technical employees of the Air District, felt sampling in representative locations using highly accurate, reference method equipment was the only reputable approach (being broadly, though not unanimously, accepted in the scientific and government communities [30]) and that alternatives would not provide trustworthy information.

Third, participants disagreed on the practical importance of fine-grained measurement. Air quality measurement is used for exposure assessment as part of the risk assessment paradigm in environmental decision-making.‡ In practice, community exposure to pollutants is often estimated using computational models and estimated values rather than fine-grained, *in situ* measurements [30]. We attended town hall meetings in which officials presented health impact assessments based entirely on modeling. On the other hand, many parties, including some representatives of regulatory agencies, were excited by the prospect of mobile sensors that could be used to collect large numbers of measurements at many locations. Some activists who had previous experience with portable measurement devices felt that the ability to measure exposure levels enabled them to challenge government and industrial assessments.

"Modeling is flawed, we measured." – *Speaker at community group meeting*

In this section, we have described various perspectives on air quality management and ways in which organizations interact with each other, particularly through data. Different perspectives on the practical goals and methods of air quality measurement complicate the collection and use of data to inform decision-making. In the next section, we discuss participants' perspectives on the specific question of introducing mobile sensing devices into this landscape.

**CRITICAL PERSPECTIVES ON CITIZEN SENSING**

Proposing mobile air quality sensing to participants, as when we described our street sweeper deployment, was provocative and sometimes controversial. Participants themselves identified many potential benefits of mobile sensing, such as learning more about which areas have the worst air quality and why, or designing and measuring the success of remediation strategies. Many saw mobile sensing as a way to free citizens from a reliance on data from government organizations, empowering them to question findings with which they disagree and investigate issues that might otherwise be ignored. However, the idea of conducting mobile sensing outside of the existing regulatory framework raised questions (and doing so inside the existing regulatory framework does not currently seem to be on the horizon). For example, while some participants supported a "street science" approach in which citizens play an active role [6], others suggested that such an approach would be unhelpful, unscientific, or even dangerous.

Below, we highlight critiques that a researcher proposing to deploy mobile sensing devices in the environmental action landscape is likely to encounter. A researcher introducing new forms of data collection or analysis must carefully consider these questions in designing their research programs and technologies and stay alert to these critiques in a variety of settings – participants' positions did not align predictably along organizational lines and the critiques were often implicitly rather than explicitly raised.

**Is Data Politically Relevant?**
*"Forget the monitoring. We're going for the solution."*

People were often very interested in mobile sensing as a means to influence other people's opinions or to pressure other people to take action. They believed that more data would help them make a more convincing case to policy makers, gain media attention which would pressure policy makers and/or corporations to take action, or galvanize (currently uninvolved) individuals from the community to "make noise" or band together to advocate for action.

However, it became clear that there are many situations in which actors in the environmental decision-making process are not interested in data. For example, many people were not particularly inclined to adopt or endorse mobile sensing for the purpose of informing *their own* beliefs. Many of our participants (from government officials to citizens) had already formed strong beliefs about a given environmental situation, and they did not anticipate that more data would strongly impact their opinion. Some of these beliefs were informed by data collection activities that had already occurred, but often these beliefs did not rely on *in situ* data and were instead the result of one or more of the following thought processes: (1) *scientific interpretation* of a situation based on known principles (e.g., housing within 1000 feet of a highway is likely to be exposed to hazardous levels of emissions from vehicles); (2) *regulatory interpretation* of a situation based on the expectation that failure to follow environmental regulations may create a hazard (e.g., trucks idling for longer periods than the regulations allow are likely to be emitting hazardous levels of pollutants); and (3) *personal interpretation* of a situation based on the experiences of the population living near a potential hazard (e.g., individuals in the neighborhood are sick and their illness is likely due to emissions from the steel plant).

---

‡ The dominant paradigm focuses on risk: *risk assessment* to estimate the magnitude and probability of incurring some loss or cost in the face of a hazard, and *risk management* to select among alternative responses [31]. Risk assessment comprises the steps of *hazard identification* (determining whether exposure causes health effects), *dose-response assessment* (relating the magnitude of exposure to health effects), *exposure assessment* (estimating how often and at what concentrations humans are exposed to the hazard), and a summary *risk characterization* (estimating the overall probability and severity of health effects) [31]. There are other important paradigms, such as those based on the precautionary principle [34] and alternatives assessment [27]. However, with its established position in the status quo and its close coupling to the scientific method, risk assessment *at present* remains at the core of policy-making in the U.S. and the E.U. [35] – particularly for air quality management [30].

In some cases, people explained that data would not contribute to their cause because the process had reached a stage or relied on a strategy where additional data would be irrelevant. For example, some community members proposed that a local factory simply replace all toxic materials with non-toxic substitutes, thereby eliminating the need to measure exposure levels. As another example, we sometimes heard that a given problem had become a well-established fact and additional data would not contribute to the discussion because the focus had shifted to remediation.

> "At that point he said, 'Forget the monitoring. We're going for the solution.' They know it's bad here. They've done the monitoring." – *Concerned citizen and activist*

These situation-based objections do not mean that mobile sensing is inapplicable in all situations. Researchers should be prepared for – but not immediately discouraged by – responses that "more data" is unwanted; a particular type of data may in fact be relevant to a different stage of a given campaign, or to a different campaign entirely.

**Is the Data "Good Enough" to be Usable?**
*"You just can't use that sort of data."*

The Air District uses expensive, high quality equipment that is carefully maintained and operated by trained staff and is audited by other agencies to ensure accuracy. By contrast, mobile environmental sensing builds on sensing technologies that are cheaper and require less maintenance and expertise to operate. By themselves, these lower-end sensors will generally be less accurate and precise, implying that more samples and more sophisticated statistical techniques will be required to produce good results [17]. While most participants recognized this as a limitation, they also generally seemed to appreciate that less accurate data could be valuable for appropriate purposes.

> "The rules that [the Air District] had to follow...I always think that they're too conservative...more data than the couple of points that they have has certainly been my idea...Some people will say, 'You just can't use that sort of data.' After a certain point, you can say, 'No, we can use it for this type of thing.'" – *Air quality consultant*

For example, participants (including some from official data collection organizations) expressed that less accurate but more lightweight data collection methods could be useful for determining high-order effects in local variation (e.g., identifying "hot spots"). A few also observed that sufficient quantities of data would overcome the precision loss associated with cheaper instruments.

**Can Citizens Participate Credibly in Environmental Sensing?**
*"The public doesn't know how to handle complex equipment."*

Various issues were raised in regards to the public's ability to conduct responsible science. Some people believe that community groups and individuals can meaningfully participate in carefully designed interventions with appropriate roles and training.

> "A lot of people who don't have engineering degrees, don't have scientific degrees, when it's affecting their neighborhood, they get educated pretty quickly." – *Planner and NGO volunteer*

However, some people question the general public's qualifications to collect or interpret data. Data collected by community groups tends to be dismissed if they can not prove that their methods are credible. One group talked about how their first tests, using off-the-shelf devices from Home Depot, were dismissed as "not good science." They had recently conducted a study using more rigorous methods and professional equipment, and their presentation slide describing this newer study proclaimed, "The testing involved good science." On a related note, representatives of air quality districts in multiple jurisdictions said that environmental groups frequently ask to borrow monitoring equipment. One representative said they explain to these groups that the equipment is sophisticated, expensive, bulky, and requires specific methodologies to yield accurate results. Another representative described a recent project with the community where the Air District had to do a lot of "hand-holding" to ensure the quality of the results.

> "The public doesn't know how to handle complex equipment." – *Data collection agency employee*

Not surprisingly, the various parties often expressed suspicions of each other. Citizens often questioned the motives and methods of government organizations and corporations; conversely, some participants worried that citizen groups might inadvertently or even intentionally distort data to prove a point, or that the public might overreact to isolated values or inaccurate data. One person proposed that non-experts should be shown only high-level results such as "safe" and "unsafe" rather than being shown the actual data values being collected on their devices. These issues speak to the importance of designing sensing tools that establish credibility, such as devices that do not require high levels of expertise to yield accurate results and mechanisms with which users can authenticate data.

In this section, we have seen how the idea of mobile participatory sensing – and of using user-collected data in general – provoked definitive and often quite sophisticated critiques from all parties. Having discussed the perceived relevance of mobile sensor data for environmental action, we now turn to practical implications for action and design.

**FROM DATA TO PRACTICAL ACTION**
In this section, we connect our findings with pragmatic considerations for researchers working in the area of environmental action, and we discuss implications for HCI research agendas in this area.

**The Framing of Academic Research is Suspect**
As we have seen, the base belief of many of our participants is that personal or organizational action is more significant than new health data, environmental data, or technology, so researchers must prepare for blunt skepticism about (1) the value of academic research and scientific knowledge and (2) the potential of new technology to make a difference. We contrast this with the more common experience of HCI researchers, in which potential users and study participants are recruited from those who are already supportive of technology and academic research, perceive potential benefits, or are compensated. This common perception that participation in research is a waste of time – "but you're not doing anything *for us*" – on the part of grassroots environmental justice activists and concerned citizens is well documented in environmental activism research [5].

> "We just get studied and studied and studied, and stuff on the street just remains the same, nothing ever changes. A thousand PhDs have come through here, you know. The people write their thesis and get their degree, and no one ever sees them again and we don't know where their data went to and we don't know what the report was about, and nothing on the street changed, right?" – *Community group leader*

However, we were less prepared for the amount of skepticism from employees of government regulatory agencies and large environmental action organizations, since they often had substantial scientific/technical background. They, too, reported experiences with researchers (e.g., scientists developing new instruments or conducting studies) who were perceived as promising a great deal, requiring significant effort to educate and support, and ultimately delivering nothing of relevance.

We learned several useful strategies from our experiences. First, having robust, deployable technological artifacts or other signs that the research is oriented toward action can help build credibility with most actors. Second, explicit and candid discussions of *quid pro quo* can be helpful. Third, local activist organizations expect (and even require) that they will play the role of an intermediary between researchers and community members, and this facilitation can be critical for lending credibility to researchers, coaching researchers in appropriate language and presentation methods, and keeping community members' comments focused on relevant issues.

**Data Must Be Made Relevant to Action**

Given the consistent focus on social and political action expressed by the participants, it is clear that data must be presented in a way that connects directly to such action. We have previously proposed [17,28] that mobile sensing can help move toward a model in which lay persons can participate in environmental action by collecting and engaging with air quality data. This proposal can be seen as following decades of action research (such as community-based participatory research in the health sciences, in which research plans and goals are co-developed with the population under study [25]) and speaks to a growing participatory trend in environment policy and environmental action research (whether under the name citizen science [19], street science [6], or democratizing science [23]). However, as one would also expect, simply providing data in some form is not enough. Significant effort and expertise (technical, political, etc.) are required to translate raw data into implemented solutions.

> "Just about every agency has tools now and they all tell you, 'Well, just go to our website.' 'What do we do about this problem?' You know, 'Just go to our website.' And everyone has some kind of tool you can use to figure out what's horrible in your community and then it's our challenge to figure out what to do with that information. How do we convert that into action?" – *Community group leader*

This has direct implications for data visualizations and interfaces such as that shown in Figure 3. Participants were critical of representations that did not directly imply action, but rather simply raised awareness or satisfied curiosity. Systems may be most effective in the environmental action context if they provide a unified interface for exploring data and taking actions; for example, some participants suggested that visualizations should include mechanisms for communicating with policy makers.

> "[This visualization shows] this little continuum up here from good to bad and that helps a little bit, but… where do we go from there?" – *Community group volunteer*

A corollary is that decisions to include certain data in an interface (whether for concerned citizens, analysts, or decision-makers) should consider factors that affect whether it will be perceived as actionable or worthy of being actionable. First, personal views on health impacts affect the perception of what data is relevant. For example, it is straightforward to find small sensors for carbon monoxide and other EPA criteria gases, so these are what mobile sensing researchers and artists measure – but these gases are not considered to be a key problem by those activists whose focus issue is airborne particulate matter. Second, received opinions on current science affect perception of relevance. For example, because of the way that regulatory measurements are taken, participants viewed certain air pollutants to be "regional" (having the same concentration over large areas) and therefore uninteresting to measure – even though studies in fact point to the existence of variation at street level. Finally, pragmatism comes into play, since some measurements are seen as more actionable than others (e.g., when the cause of an emission can be easily localized, identified, and addressed). Again, researchers need to appreciate that a given technology may not be relevant to all situations and that they should consider focusing interventions on campaigns (or stages within campaigns) where their data has the most potential to connect to action.

A final consideration is that meaningful analysis and effective presentation may require access to additional technical resources, such as planning databases (e.g., cross-referenced data on local polluters, traffic patterns, weather models, epidemiological data, housing costs, etc.) and tools (e.g., geographic information system (GIS) software commonly used by urban planners to present results [6]). Lack of access to these systems can result in failure to connect to decision-making processes, so researchers should be aware that their partners may not have easy access to these resources. There are opportunities for researchers to develop integrated tools and integration toolkits to facilitate access to these resources, as well as to develop collaborative features to connect advocacy groups with relevant technical experts for assistance with their use.

**Design to Span Organizations**

While it is tempting to approach these problems in view of an ideal user-centered design process – e.g., designing a tool for a specific body of users such as an activist group – the problems here do not always lend themselves to this. As discussed in the previous section, a particular activity to which data is relevant (e.g., a campaign to reduce emissions from a specific plant) may involve coalitions of disparate member groups and is likely to outlast some of these groups. Our participants described functions that span groups but are poorly supported by current technologies. Hence, a key design implication is to develop tools that meet the needs of coalitions of groups without necessarily being central to the daily needs of any given group.

One concept described by participants was to provide social networking tools specialized for (advocacy) groups rather than for individuals. One participant noted the need to find other groups that are working on related problems, a "social mapping of the organizations available that are working on issues in the neighborhood." Tools to map the organizational, geographical and topical landscape surrounding one's own group would facilitate the critical coalition "hitching" strategy mentioned in an earlier section; grassroots organizations need ways to link to national and global agendas, and the appropriate entry points to the larger environmental organizations with those agendas may lie outside of members' individual social networks. Another participant observed that it would be useful for visualizations to link to air quality education, advocacy groups, and local issues ("What special things are happening right now that are needing attention"). A natural extension of these proposals would be a tool that integrated views of data (including environmental, epidemiological, or health indicator data, thus providing insight into the problem) with social software (facilitating connections between groups once insights were found).

From our fieldwork, we identified an additional design opportunity: that of designing tools to support long-lived campaigns conducted by coalitions of ephemeral groups. Several participants noted the general issue that "there are groups that are forming and groups that are going out of existence all the time." Continuity in monitoring is critical to ensure accountability: collectively, organizations involved in a campaign may need to verify that remediation actions are effective and continue to be applied. A particular group may exhaust itself after having reached a certain point, but the campaign in a community may continue for years or decades. Therefore, the advocacy network would benefit greatly from collaborative work tools designed to support persistence and knowledge transfer across individuals and organizations, in order to provide continuity through all stages of a campaign (data collection, analysis, political communications, monitoring, etc.) and for its entire duration.

**CONCLUSIONS**

In this paper, we have presented a qualitative analysis of the landscape of environmental action. In so doing, we have sought to illuminate the context in which technology will be received (or rejected) and to identify key insights that will help researchers frame meaningful interventions. Specifically, we have discussed: (1) the landscape of environmental action, and how researchers can orient themselves in this setting; (2) key critiques of data interventions, and how researchers can design interventions and interact with communities to mitigate these critiques; and (3) practical issues in using technological innovation to improve the quality of environmental decision-making, and the implications of these issues for ICT agendas for environmental action. We have also called out several design opportunities for contributing to environmental action, such as the need for social mapping tools to connect advocacy groups, the need for collaborative tools that span advocacy groups in order to preserve continuity during lengthy campaigns that outlive individual groups, and the need for trust mechanisms that establish the credibility of data collected by non-expert users. In our own work, as we move from our initial learning experience with the street sweepers to deployments of a personal sensing device, we expect these lessons will continue to help us navigate the landscape of environmental community action.

**ACKNOWLEDGEMENTS**

We gratefully acknowledge the support of the City of San Francisco Office of the Mayor, Dept. of Public Works, and Dept. of the Environment. We are grateful to Ron Cohen, Liz Goodman, Ben Hooker, Reza Naima, Paul Wooldridge, the anonymous reviewers, and the participants for their valuable contributions. Common Sense builds in part on the Participatory Urbanism [28] and N-SMARTS [17] projects.